# Invariant quantities of a nondepolarizing Mueller matrix

José J. Gil[1*], Ignacio San José[2]

[1]*Universidad de Zaragoza. Pedro Cerbuna 12, 50009 Zaragoza, Spain.*
[2]*Instituto Aragonés de Estadística. Gobierno de Aragón. Bernardino Ramazzini 5, 50015 Zaragoza, Spain.*
*Corresponding author: ppgil@unizar.es*

Orthogonal Mueller matrices can be considered either as corresponding to retarders or to generalized transformations of the polarization basis for the representation of Stokes vectors, so that they constitute the only type of Mueller matrices that preserve the degree of polarization and the intensity of any partially-polarized input Stokes vector. The physical quantities which remain invariant when a nondepolarizing Mueller matrix is transformed through its product by different types of orthogonal Mueller matrices are identified and interpreted, providing a better knowledge of the information contained in a nondepolarizing Mueller matrix.

## 1. Introduction

The wide scope of applications of Mueller polarimetry justifies the necessity of the best knowledge of the physical information contained in a given Mueller matrix $\mathbf{M}$. In fact, the intricate mathematical structure of $\mathbf{M}$ makes it difficult to identify and interpret such information. While serial [1-5] and parallel [6-10] decompositions of $\mathbf{M}$ constitute relevant approaches to achieve this goal, interesting parameterizations of a general Mueller matrix $\mathbf{M}$ have also been obtained recently from the identification of physical parameters that are invariant under retarder transformations [11].

The physical significance of this kind of *reversible* transformations relies on the fact that the most general lossless and nondepolarizing transformation of a Mueller matrix $\mathbf{M}$ is given by [12]

$$\mathbf{M}_{R2}\mathbf{M}\mathbf{M}_{R1}, \qquad (1)$$

where $\mathbf{M}_{R1}$ and $\mathbf{M}_{R2}$ are orthogonal Mueller matrices representing respective retarders (recall that the serial combination of a number of retarders is equivalent to a retarder).

*Nondepolarizing* (or *pure*) Mueller matrices, being a particular class of Mueller matrices, play a very important role in polarimetry because 1) they represent the more basic linear polarimetric interactions, 2) any Mueller matrix can be considered as an ensemble average of nondepolarizing Mueller matrices [13-15], 3) many media behave, at least in the first-order approximation, like nondepolarizing, so that nondepolarizing Mueller matrices represent, by themselves, a wide range of polarimetric behaviors [16] and 4) any depolarizing Mueller matrix has an associated *reference pure Mueller matrix* [17].

Moreover, pure Mueller matrices have a very peculiar mathematical structure and consequently, contrary to what should be expected, the identification and interpretation of complete and meaningful sets of their invariant quantities under serial combinations of the corresponding media with retarders are not straightforwardly derived from the general results presented in Ref. [11]. Note, for instance, that while the arrow form $\mathbf{M}_A$ [5] of a general Mueller matrix $\mathbf{M}$ depends on ten scalar quantities, it involves only two scalar independent parameters when $\mathbf{M}$ corresponds to a nondepolarizing medium. In addition, quantities like $\mathrm{tr}\mathbf{M}$ and $\det\mathbf{M}$ have specific interpretations when $\mathbf{M}$ is a pure Mueller matrix.

Beyond the main objective of this work that consist of identifying and interpreting the invariant quantities of a pure Mueller matrix under different types of reversible transformations, it is worth to mention that such approaches can be applied for endoscopic Jones polarimetry when it is focused on the measurement of specific invariant properties and, therefore, where certain properties of the fiber devices used in the polarization state generator and polarization state analyzer (PSG and PSA respectively) do not necessarily have to be predetermined (for instance the effective twist of the optical fiber).

Additional applications can be found in the analysis of material samples constituted by (or polarimetrically equivalent to) serial combinations of retarders and nondepolarizing media with non-negligible diattenuation, which are encountered in a wide variety of practical situations as, for instance, birefringent layered structures (including biological tissues) [18], LCD devices [19], etc.

For the sake of clarity and self-consistency of this work, let us recall that *dual retarder transformations* $\mathbf{M}_{R2}\mathbf{M}\mathbf{M}_{R1}$ can be interpreted as changes of generalized reference frames (elliptical in general) for the representation of the input and output states of polarization, which preserve important properties like the intensity, the degree of polarization and the coherence properties (i.e., the electromagnetic degree of coherence [20], the intrinsic degrees of coherence [21], or the Luis's overall degree of coherence [22]).

The subclass of *single retarder transformations* can be interpreted as that constituted by dual retarder transformations with a common modification for the input and output generalized bases $\left(\mathbf{M}_{R2}=\mathbf{M}_{R1}^{T}\right)$.

*Dual rotation transformations* $\mathbf{M}_{RC2}\mathbf{M}\mathbf{M}_{RC1}$ can be interpreted as produced by independent rotations of the Cartesian reference frames for the representation of the input and output states of polarization, that is, both $\mathbf{M}_{RC1}$ and $\mathbf{M}_{RC2}$ have the form of rotation matrices (or circular retarders, see Section 3).

*Single retarder transformations* is a subclass of reversible transformations where both input and output Cartesian reference frames are rotated jointly $\left(\mathbf{M}_{RC2}=\mathbf{M}_{RC1}^{T}\right)$.

To address the required specific stand-alone mathematical study and physical interpretation of each type of reversible transformation of a pure

Mueller matrix Section 2 is devoted to a summary of the necessary theoretical background on the structure of pure Mueller matrices. Section 3 deals with the representation of changes of the reference frames for both incoming and outgoing polarization states of electromagnetic waves that interact with the medium under consideration; as indicated above, such transformations are represented mathematically by means of orthogonal Mueller matrices, which, in turn, have the form of Mueller matrices associated with retarders (birefringent media). The general form of dual retarder transformations is considered in Section 4, where the corresponding invariant quantities are identified and interpreted. Sections 5 to 7 deal with the mathematical formulation and physical interpretation of respective particular forms of dual retarder transformations. The main results are summarized and discussed briefly in Section 8.

## 2. Pure Mueller matrices and their structure

To simplify, and to make easier the interpretation of certain operations with Mueller matrices, when appropriate we will use the following partitioned block expression of a Mueller matrix $\mathbf{M}$, which reflects important features of the structure of the information contained in $\mathbf{M}$ [17]

$$\mathbf{M} = m_{00}\hat{\mathbf{M}}$$

$$\hat{\mathbf{M}} \equiv \begin{pmatrix} 1 & \mathbf{D}^T \\ \mathbf{P} & \mathbf{m} \end{pmatrix} \qquad \mathbf{m} \equiv \frac{1}{m_{00}}\begin{pmatrix} m_{11} & m_{12} & m_{13} \\ m_{21} & m_{22} & m_{23} \\ m_{31} & m_{32} & m_{33} \end{pmatrix}$$

$$\mathbf{D} \equiv \frac{1}{m_{00}}(m_{01}, m_{02}, m_{03})^T \qquad \mathbf{P} \equiv \frac{1}{m_{00}}(m_{10}, m_{20}, m_{30})^T \qquad (2)$$

where $m_{00}$ is the *mean intensity coefficient* (i.e., the *transmittance*, or *gain* of $\mathbf{M}$ for input unpolarized states), $\hat{\mathbf{M}}$ is the *normalized Mueller matrix*, and $\mathbf{D}$ and $\mathbf{P}$ are respectively the *diattenuation vector* and the *polarizance vector* of $\mathbf{M}$ [2], whose absolute values are the *diattenuation*, $D \equiv |\mathbf{D}|$ and the *polarizance*, $P \equiv |\mathbf{P}|$.

Recall that polarizance and diattenuation share a common physical nature because $D$ is both the diattenuation of $\mathbf{M}$ and the polarizance of the *reverse Mueller matrix* [24,25]

$$\mathbf{M}^r \equiv \mathrm{diag}(1,1,-1,1)\,\mathbf{M}^T\,\mathrm{diag}(1,1,-1,1) \qquad (3)$$

corresponding to the same interaction as $\mathbf{M}$, but interchanging the input and output directions ($\mathbf{M}^T$ being the transposed matrix of $\mathbf{M}$). Analogously, $P$ is both the polarizance of $\mathbf{M}$ and the diattenuation of $\mathbf{M}^r$.

In accordance with the common convention, the term *nondepolarizing*, or *polarimetrically pure*, is used here to refer to media that preserve the degree of polarization of input states of polarization in both *forward* and *reverse* directions [26]. A proper measure of the closeness of $\mathbf{M}$ to a pure Mueller matrix is given by the depolarization index [27] (also called degree of polarimetric purity)

$$P_\Delta \equiv \sqrt{\frac{D^2 + P^2 + 3P_S^2}{3}} \qquad (4)$$

where the *degree of spherical purity* $P_S$ defined as [28]

$$P_S = \|\mathbf{m}\|_2 / \sqrt{3} \qquad (5)$$

$\|\mathbf{m}\|_2$ being the Frobenius norm of the 3×3 submatrix $\mathbf{m}$. For its part, $P_S$ constitutes a measure of the closeness of $\mathbf{M}$ to the Mueller matrix $\mathbf{M}_R$ of a retarder [29] (regardless of the value of the retardance, see below)

In the case of a pure Mueller matrix (also called Mueller-Jones matrix) $\mathbf{M}_J$, it satisfies $P_\Delta(\mathbf{M}_J) = 1$ [27] and $P(\mathbf{M}_J) = D(\mathbf{M}_J)$ [30], so that

$$2D^2 + 3P_S^2 = 3 \qquad (6)$$

and therefore, $D$ fully determines the sharing of polarimetric purity of a nondepolarizing medium between its polarizance-diattenuation $D$ and its spherical purity $P_S$ $(P_S^2 = 1 - 2D^2/3)$.

The general form of the Mueller matrix of a retarder is

$$\mathbf{M}_R = \begin{pmatrix} 1 & \mathbf{0}^T \\ \mathbf{0} & \mathbf{m}_R \end{pmatrix} \qquad \begin{bmatrix} \det \mathbf{m}_R = +1 \\ \mathbf{m}_R^T = \mathbf{m}_R^{-1} \end{bmatrix} \qquad (7)$$

which has two eigenstates (i.e., two Stokes eigenvectors, represented by two antipodal points in the Poincaré sphere)

$$\hat{\mathbf{s}}_{R+} \equiv \begin{pmatrix} 1 \\ \mathbf{u}_R \end{pmatrix} \qquad \hat{\mathbf{s}}_{R-} \equiv \begin{pmatrix} 1 \\ -\mathbf{u}_R \end{pmatrix} \qquad (8)$$

so that the input Stokes vectors are transformed by $\mathbf{M}_R$ through a rotation of an angle $\Delta$ (retardance) about the axis defined by the unit Poincaré vector $\mathbf{u}_R$

$$\mathbf{u}_R \equiv \begin{pmatrix} \cos 2\chi \cos 2\varphi \\ \cos 2\chi \sin 2\varphi \\ \sin 2\chi \end{pmatrix} \qquad (9)$$

where $\varphi$ and $\chi$ are the respective azimuth and ellipticity angle of the polarization ellipse of the fast eigenstate $\hat{\mathbf{s}}_{R+}$ of $\mathbf{M}_R$.

Thus, the retarder can also be represented in the Poincaré sphere by the corresponding *straight retardance vector* [29]

$$\overline{\mathbf{R}} \equiv \frac{\mathbf{R}}{\pi} = \frac{\Delta}{\pi}\mathbf{u}_R \qquad (10)$$

where $\mathbf{R}$ is the *retardance vector* [2] (whose absolute value $\Delta$ is in the interval $0 \leq \Delta \leq \pi$). When $\chi = 0$ the retarder is said to be *linear* ($\overline{\mathbf{R}}$ lies in the equatorial plane of the Poincaré sphere); when $\chi = \pm\pi/2$ the retarder is said to be right- or left-*circular* ($\overline{\mathbf{R}}$ lies in the vertical axis of the Poincaré sphere), the corresponding Mueller matrix having the mathematical form of a rotation matrix determined by the single parameter $\Delta$ (see Section 3), and when $\chi \neq 0, \pm\pi/2$ the retarder is said to be *elliptic*.

Note that orthogonal 4×4 matrices with forms other than $\mathbf{M}_R$ in Eq. (7), are not Mueller matrices [7]. Let us also recall that the inverse matrix $\mathbf{M}_R^{-1} = \mathbf{M}_R^T$ of a retarder $\mathbf{M}_R$ represents another retarder with the same eigenstates as $\mathbf{M}_R$ but exhibiting opposite retardance (−$\Delta$).

Another type of pure Mueller matrix, with complementary properties to those of $\mathbf{M}_R$, is that of a *normal* [31,32] (or *homogeneous* [33]) diattenuator [2]

$$\mathbf{M}_D \equiv m_{00}\begin{pmatrix} 1 & \mathbf{D}^T \\ \mathbf{D} & \mathbf{m}_D \end{pmatrix}$$

$$\mathbf{m}_D \equiv \sin\kappa\,\mathbf{I}_3 + (1 - \sin\kappa)\hat{\mathbf{D}} \otimes \hat{\mathbf{D}}^T$$

$$\left(\sin\kappa \equiv \sqrt{1 - D^2} \quad \mathbf{I}_3 \equiv \mathrm{diag}(1,1,1) \quad \hat{\mathbf{D}} \equiv \mathbf{D}/D\right) \qquad (11)$$

where $\sin\kappa$ (with $0 \leq \kappa \leq \pi/2$) is the *counterpolarizance* [29] (or *counterdiattenuation*) of $\mathbf{M}_D$.

The eigenstates of $\mathbf{M}_D$ are given by

$$\hat{\mathbf{s}}_{D+} \equiv \begin{pmatrix} 1 \\ \hat{\mathbf{D}} \end{pmatrix} \qquad \hat{\mathbf{s}}_{D-} \equiv \begin{pmatrix} 1 \\ -\hat{\mathbf{D}} \end{pmatrix} \qquad (12)$$

whose respective eigenvalues

$$p_1^2 \equiv m_{00}(1+D) \qquad p_2^2 \equiv m_{00}(1-D) \qquad (13)$$

determine the maximal $(p_1^2)$ and minimal $(p_2^2)$ intensity coefficients of $\mathbf{M}_D$.

When $D=1$, $\mathbf{M}_D$ is a singular matrix that corresponds to a pure normal *polarizer*. Provided $D<1$, the inverse matrix $\mathbf{M}_D^{-1}$ of $\mathbf{M}_D$ is given by

$$\mathbf{M}_D^{-1} \equiv \frac{1}{m_{00}\sin\kappa}\begin{pmatrix} 1 & -\mathbf{D}^T \\ -\mathbf{D} & \mathbf{m}_D \end{pmatrix} \qquad (14)$$

Although $\mathbf{M}_D^{-1}$ satisfies Cloude's criterion [14,34,35] required for it to be a Mueller matrix (i.e., the nonzero eigenvalue of its associated covariance matrix is positive), $\mathbf{M}_D^{-1}$ produces output polarization states whose intensity is higher than the corresponding input ones. Obviously, this feature of amplifying the intensity only occurs, in some way, in certain artificial arrangements [36], so that passivity is an additional criterion required (within the experimental error tolerance) for measured Mueller matrices [5,6,30,37-41].

Since the transformations studied in this work are formulated as products of Mueller matrices, let us recall that the Mueller matrix representing the linear polarimetric effect of a serial combination of a number of material elements interposed in the pathway of an electromagnetic wave is given by the ordered product of the Mueller matrices of the combined elements. Furthermore, given a pure Mueller matrix $\mathbf{M}_J$, it can always be expressed as the product of a retarder and a normal diattenuator (*polar decomposition*) in either of the two possible relative positions [1,2,42]

$$\mathbf{M}_J = \mathbf{M}_R \mathbf{M}_D = \mathbf{M}_P \mathbf{M}_R \qquad (15)$$

where $\mathbf{M}_D$ and $\mathbf{M}_P$, are mutually related through

$$\mathbf{M}_P = \mathbf{M}_R \mathbf{M}_D \mathbf{M}_R^T \qquad (16)$$

Therefore,

$$\mathbf{M}_J = m_{00}\begin{pmatrix} 1 & \mathbf{D}^T \\ \mathbf{m}_R \mathbf{D} & \mathbf{m}_R \mathbf{m}_D \end{pmatrix} = m_{00}\begin{pmatrix} 1 & (\mathbf{m}_R^T \mathbf{P})^T \\ \mathbf{P} & \mathbf{m}_P \mathbf{m}_R \end{pmatrix} \qquad (17)$$

with

$$\mathbf{P} = \mathbf{m}_R \mathbf{D} \qquad \mathbf{D} = \mathbf{m}_R^T \mathbf{P}$$

$$\mathbf{m} = \mathbf{m}_R \mathbf{m}_D = \mathbf{m}_P \mathbf{m}_R \qquad (18)$$

so that, by considering Eqs. (11) and (18),

$$\mathbf{m} \equiv \sin\kappa\,\mathbf{m}_R + (1-\sin\kappa)\hat{\mathbf{P}}\otimes\hat{\mathbf{D}}^T \qquad \begin{pmatrix} \hat{\mathbf{P}} \equiv \mathbf{P}/D \\ \hat{\mathbf{D}} \equiv \mathbf{D}/D \end{pmatrix} \qquad (19)$$

The polar decomposition of $\mathbf{M}_J$ does not have a symmetric serial structure and admits two distinct forms depending on the order of the equivalent retarder and the equivalent normal diattenuator. Thus, for some analyses, it is also useful to consider the following symmetric form of the general serial decomposition of $\mathbf{M}_J$ [3,43]

$$\mathbf{M}_J = m_{00}\,\mathbf{M}_{RL2}\,\mathbf{M}_{RL0}\,\hat{\mathbf{M}}_{DL0}\,\mathbf{M}_{RL0}\,\mathbf{M}_{RL1} \qquad (20)$$

where $\mathbf{M}_{RL1}$ and $\mathbf{M}_{RL2}$ are the Mueller matrices of respective *entrance* and *exit* linear retarders, with respective azimuths $(\varphi_1,\varphi_2)$ and retardances $(\Delta_1,\Delta_2)$, $\mathbf{M}_{RL0}$ represents a horizontal linear retarder with retardance $\Delta_0/2$ and $\hat{\mathbf{M}}_{DL0}$ is the normalized matrix of a horizontal linear diattenuator of the form

$$\mathbf{M}_{DL0} = m_{00}\begin{pmatrix} 1 & \mathbf{D}_0^T \\ \mathbf{D}_0 & \mathbf{m}_{DL0} \end{pmatrix} \qquad (21.a)$$

with

$$\mathbf{D}_0 \equiv (\cos\kappa,0,0)^T \qquad \mathbf{m}_{DL0} \equiv \mathrm{diag}(1,\sin\kappa,\sin\kappa)$$

$$\left(\sin\kappa \equiv \sqrt{1-D^2} \quad \cos\kappa \equiv D \quad 0\le\kappa\le\pi/2\right) \qquad (21.b)$$

and therefore

$$\mathbf{M}_J = m_{00}\begin{pmatrix} 1 & (\mathbf{m}_{LR1}^T \mathbf{m}_{LR0}^T \mathbf{D}_0)^T \\ \mathbf{m}_{RL2}\mathbf{m}_{RL0}\mathbf{D}_0 & \mathbf{m}_{RL2}\mathbf{m}_{RL0}\mathbf{m}_{DL0}\mathbf{m}_{RL0}\mathbf{m}_{RL1} \end{pmatrix} \qquad (22)$$

The above polar and symmetric decompositions of $\mathbf{M}_J$ show that retarders and normal diattenuators can be considered as building blocks of pure Mueller matrices whose main properties and differences deserve to be summarized below.

Retarders play a key role in serial decompositions of Mueller matrices because they constitute the only type of media that preserve the degree of polarization $\mathcal{P}$ of the input polarized electromagnetic waves, regardless of the value of $\mathcal{P}$. Note that this feature implies that $\mathbf{M}_R$ is a special kind of pure Mueller matrix [i.e. $P_\Delta(\mathbf{M}_R)=1$]. In addition, $\mathbf{M}_R$ preserves the intensity; in fact, $\mathbf{M}_R$ is the only type of passive Mueller matrix whose inverse matrix is a passive Mueller matrix.

Unlike retarders, diattenuators modify the degree of polarization of partially polarized input states and produce an anisotropic reduction of the intensity of the input states. Consequently, the product of a pure Mueller matrix $\mathbf{M}_J$ and a diattenuator implies changes on the polarizance-diattenuation $D$ and on $P_\Delta$. This is the reason why this work is focused on the physical quantities that are invariant under different types of dual retarder transformations $\mathbf{M}_{R2}\mathbf{M}_J\mathbf{M}_{R1}$.

## 3. Laboratory and generalized polarization bases

Mueller polarimetry is applied to the study and characterization of material samples under a great variety of interaction conditions, namely refraction, reflection, scattering, etc. in such a manner that different, and in general independent, bases for the representation of the input and output states of polarization are used. Input and output Stokes vectors are commonly defined with respect to their own Cartesian laboratory coordinate systems $X_I Y_I Z_I$ and $X_O Y_O Z_O$, where the characteristic polarization ellipses of the input and output polarization states lie in the respective planes $X_I Y_I$ and $X_O Y_O$, while the respective directions of propagation are aligned with axes $Z_I$ and $Z_O$. Arbitrary rotations of counterclockwise angles $\theta_I$ and $\theta_O$ of the input and output reference frames about the respective axes $Z_I$ and $Z_O$ involve transformations of the input and output Stokes vectors of the form

$$\mathbf{s}_I' = \mathbf{M}_G(\theta_I)\mathbf{s}_I \qquad \mathbf{s}_O' = \mathbf{M}_G(\theta_O)\mathbf{s}_O$$

$$\mathbf{M}_G(\theta_i) \equiv \begin{pmatrix} 1 & 0 & 0 & 0 \\ 0 & \cos 2\theta_i & \sin 2\theta_i & 0 \\ 0 & -\sin 2\theta_i & \cos 2\theta_i & 0 \\ 0 & 0 & 0 & 1 \end{pmatrix} \qquad (i=I,O) \qquad (23)$$

so that the Mueller matrix $\mathbf{M}_J'$ corresponding to the transformed reference frames is given by

$$\mathbf{M}_J' = \mathbf{M}_{GO}\mathbf{M}_J\mathbf{M}_{GI}^T = \mathbf{M}_G(\theta_O)\mathbf{M}_J\mathbf{M}_G(-\theta_I) \qquad (24)$$

This kind of transformation is called a *dual rotation transformation* of $\mathbf{M}_J$ [11]. In the particular (but very common) case that input and output

Cartesian reference frames are rotated jointly $(\theta_O = \theta_I \equiv \theta)$, the Mueller matrix of the corresponding *single rotation transformation* is

$$\mathbf{M}'_J = \mathbf{M}_G \mathbf{M}_J \mathbf{M}_G^T = \mathbf{M}_G(\theta) \mathbf{M}_J \mathbf{M}_G(-\theta) \tag{25}$$

It is worth recalling that matrix $\mathbf{M}_G(\theta)$ has the mathematical form of the Mueller matrix of a circular retarder exhibiting retardance $\Delta = 2\theta$ [44], and therefore, rotation transformations can be formulated as serial combinations where $\mathbf{M}_J$ is sandwiched by circular retarders.

Although, as indicated above, Stokes vectors are commonly defined with respect to Cartesian reference frames *XYZ*, it is also possible to define non-Cartesian elliptical or circular reference bases for its representation. Leaving aside the fact that such non-Cartesian reference systems cannot be physically realized through rotations of *XYZ*, they are mathematically defined as

$$\mathbf{s}'_I = \mathbf{M}_{RI} \mathbf{s}_I \qquad \mathbf{s}'_O = \mathbf{M}_{RO} \mathbf{s}_O \tag{26}$$

where $\mathbf{M}_{RI}$ and $\mathbf{M}_{RO}$ are orthogonal Mueller matrices (hence with the form of Mueller matrices of retarders). Thus, the transformed Mueller matrix under such *dual retarder transformation* [11] is given by

$$\mathbf{M}'_J = \mathbf{M}_{RO} \mathbf{M}_J \mathbf{M}_{RI}^T \tag{27}$$

which can also be mathematically interpreted as a serial combination where the medium corresponding to $\mathbf{M}_J$ is sandwiched by retarders (in general elliptic). In the particular case that the bases for the representation of input and output polarization states are modified jointly through a common generalized transformation, the Mueller matrix with respect to the new generalized reference frame is given by the *single retarder transformation*

$$\mathbf{M}'_J = \mathbf{M}_R \mathbf{M}_J \mathbf{M}_R^T \tag{28}$$

Obviously, since $\mathbf{M}_G$ has the form of a particular type of retarder, a dual rotation transformation is a particular type of dual retarder transformation [11], and a single rotation transformation is a particular type of single retarder transformation.

Next sections are devoted to specific analyses of the properties of pure Mueller matrices on the basis of the dual retarder transformation and the above mentioned particular cases.

## 4. Dual retarder transformation

Observe that, from Eq. (20), $\mathbf{M}_J$ can be expressed as the serial product

$$\mathbf{M}_J = \mathbf{M}_{RO} \left( m_{00} \hat{\mathbf{M}}_{DL0} \right) \mathbf{M}_{RI} \tag{29}$$

of the horizontal linear diattenuator $\mathbf{M}_{DL0} = m_{00} \hat{\mathbf{M}}_{DL0}$ sandwiched by the retarders $\mathbf{M}_{RO} \equiv \mathbf{M}_{RL2} \mathbf{M}_{RL0}$ and $\mathbf{M}_{RI} \equiv \mathbf{M}_{RL0} \mathbf{M}_{RL1}$.

Furthermore, since the serial composition of retarders is equivalent to a retarder, the quantities $m_{00}$ and $D$ are invariant under *dual retarder transformations* $\mathbf{M}_{R2} \mathbf{M}_J \mathbf{M}_{R1}$ of $\mathbf{M}_J$, where $\mathbf{M}_{R1}$ and $\mathbf{M}_{R2}$ represent arbitrary retarders.

This implies that the mean intensity coefficient $m_{00}$, together with the polarizance-diattenuation $D$, which provide equivalent information to that given by the maximal and minimal intensity coefficients

$$p_1^2(\mathbf{M}_J) = m_{00}(1+D) \qquad p_2^2(\mathbf{M}_J) = m_{00}(1-D) \tag{30}$$

constitute a complete set of independent quantities that remain unchanged when the medium represented by $\mathbf{M}_J$ is submitted to arbitrary serial combinations with retarders.

Thus, in accordance with Eq. (20) $\mathbf{M}_J$ a complete parameterization of $\mathbf{M}_J$ is given by the seven parameters

$$m_{00}, D, \Delta_0, \varphi_1, \Delta_1, \varphi_2, \Delta_2 \tag{31}$$

where $(m_{00}, D)$ determine the central horizontal diattenuator $\mathbf{M}_{DL0}$ (which accumulates the invariant information); $(\Delta_0, \varphi_1, \Delta_1)$ determine the *entrance retarder*

$$\mathbf{M}_{RI} \equiv \mathbf{M}_{RL0}(\Delta_0/2) \mathbf{M}_{RL1}(\varphi_1, \Delta_1) \tag{32}$$

and $(\Delta_0, \varphi_2, \Delta_2)$ determine the *exit retarder*

$$\mathbf{M}_{RO} \equiv \mathbf{M}_{RL2}(\varphi_2, \Delta_2) \mathbf{M}_{RL0}(\Delta_0/2) \tag{33}$$

Obviously, as it is well-known [1], $\mathbf{M}_J$ can also be parameterized through either of the two alternative polar decompositions, but the above *symmetric parameterization* has the advantage of having a unique form that, furthermore, matches with the parameterization of general Mueller matrices (depolarizing or not) based on its *arrow decomposition* [5].

## 5. Single retarder transformation

Since $\mathbf{M}_J$ depends on seven parameters and a general retarder $\mathbf{M}_R$ depends on three parameters, a total of four physical quantities of $\mathbf{M}_J$ remain invariant under single retarder transformations, two of which being $m_{00}$ and $D$ (as seen in Section 4). In general, the entrance retarder $\mathbf{M}_{RI} \equiv \mathbf{M}_{RL0} \mathbf{M}_{RL1}$ of the general serial decomposition of $\mathbf{M}_J$ in Eq. (20) is different from $\mathbf{M}_{RO}^T = \mathbf{M}_{RL0}^T \mathbf{M}_{RL2}^T$ and therefore, such decomposition does not lead straightforwardly to the identification of the additional pair of invariant quantities. Nevertheless, they can be derived from the fact that $\mathbf{P}^T \mathbf{m}^T \mathbf{D}$ and $\mathrm{tr}\mathbf{M}$ always remain invariant under a single retarder transformation of a Mueller matrix $\mathbf{M}$ [11] and, by considering Eqs. (18) and (19),

$$\mathbf{P}^T \mathbf{m}^T \mathbf{D} = \sin\kappa \, \cos^2\kappa \, \cos^2\eta + (1-\sin\kappa)\cos\eta$$
$$\mathrm{tr}\mathbf{M}_J = \sin\kappa(1+2\cos\Delta) + (1-\sin\kappa)\cos\eta \tag{34}$$

which implies that, besides $\sin\kappa$ (which only depends on $D$), the angle $\eta$ subtended between $\mathbf{P}$ and $\mathbf{D}$, together with the retardance parameter $\Delta$ of the equivalent retarder $\mathbf{M}_R$ in the polar decomposition of $\mathbf{M}_J$, necessarily remain invariant. In summary, a meaningful set of invariant quantities of $\mathbf{M}_J$ under single retarder transformations is the one constituted by

$$m_{00}, D, \eta, \Delta \tag{35}$$

It should be noted that, as expected, all these quantities are independent of the relative order of the retarder and the diattenuator in the polar decomposition of $\mathbf{M}_J$.

## 6. Dual rotation transformation

Since the rotation Mueller matrix in Eq. (23) depends on a single angular parameter $\theta$, five is the number of quantities of $\mathbf{M}_J$ that remain invariant under dual rotation transformations [see Eq. (24)]. From the analysis performed in Ref. [11] for general Mueller matrices, it turns out that the linear and circular components

$$D_L \equiv \sqrt{D_1^2 + D_2^2} \qquad D_C \equiv D_3$$
$$P_L \equiv \sqrt{P_1^2 + P_2^2} \qquad P_C \equiv P_3 \tag{36}$$

of vectors $\mathbf{D} \equiv (D_1, D_2, D_3)^T$ and $\mathbf{P} \equiv (P_1, P_2, P_3)^T$ necessarily remain invariant. Furthermore, it is straightforward to prove that the linear and circular components

$$\bar{R}_L \equiv \sqrt{\bar{R}_1^2 + \bar{R}_2^2} \qquad \bar{R}_C \equiv \bar{R}_3 \tag{37}$$

of the straight retardance vector $\bar{\mathbf{R}} \equiv (\bar{R}_1, \bar{R}_2, \bar{R}_3)^T$ also remain invariant. Therefore, besides $D$, other derived quantities like the retardance $\Delta = \pi\sqrt{\bar{R}_L^2 + \bar{R}_C^2}$, also remain invariant.

Consequently, a complete set of five independent and invariant quantities is the one constituted by

$$m_{00}, D, D_C, P_C, \bar{R}_C \quad (38)$$

(note that any other invariant quantity can be calculated from this set).

## 7. Single rotation transformation

The joint rotation of the input and output reference frames by a counterclockwise angle $\theta$ leads to the transformation of $\mathbf{M}_J$ in the form $\mathbf{M}_G(\theta)\mathbf{M}_J\mathbf{M}_G(-\theta)$ [see Eq.(25)]. Since this is a particular case of both dual rotation transformation, and single retarder transformation, both sets indicated in Eq. (35) and Eq. (38), as well as the corresponding derived quantities, remain invariant. In particular, a complete set of six independent invariants is given by

$$m_{00}, D, D_C, P_C, \bar{R}_C, \eta \quad (39)$$

showing that, as expected, and unlike in dual rotation transformations, the angle $\eta$ subtended between vectors $\mathbf{P}$ and $\mathbf{D}$ is preserved in single rotation transformations.

## 8. Conclusion

The physical information contained in a nondepolarizing (or *pure*) Mueller matrix $\mathbf{M}_J$ has been analyzed by means of the quantities that remain invariant under different types of transformations of the form $\mathbf{M}_{R2}\mathbf{M}_J\mathbf{M}_{R1}$ (*dual retarder transformations*) where the orthogonal Mueller matrices $\mathbf{M}_{R1}$ and $\mathbf{M}_{R2}$ can be interpreted either as corresponding to respective retarders or to respective changes of generalized reference frames for the input and output polarization states.

The results obtained are summarized in Table 1, where it is taken into account that any retarder represented by a given orthogonal Mueller matrix $\mathbf{M}_R$ is also fully determined by its corresponding straight retardance vector $\bar{\mathbf{R}}$.

It is remarkable that the retardance $\Delta$ of the equivalent retarder in the polar decomposition of $\mathbf{M}_J$ is preserved except for the most general case where $\mathbf{M}_{R1}$ or $\mathbf{M}_{R2}$ is elliptic, with $\mathbf{M}_{R2}^T \neq \mathbf{M}_{R1}$ ($\bar{\mathbf{R}}_2 \neq -\bar{\mathbf{R}}_1$).

Furthermore, as expected, the angle $\eta$ subtended between diattenuation and polarizance vectors $\mathbf{D}$ and $\mathbf{P}$ is preserved if and only if $\mathbf{M}_{R2}^T = \mathbf{M}_{R1}$ ($\bar{\mathbf{R}}_2 = -\bar{\mathbf{R}}_1$), while the circular and linear components of $\mathbf{D}$, $\mathbf{P}$ and $\bar{\mathbf{R}}$ are invariant under any kind of rotation transformation.

**Table 1**. Invariant quantities under different types of transformations

| Transformation | Retardance vectors | Independent invariant quantities | Other invariant quantities |
|---|---|---|---|
| Dual retarder | $\bar{\mathbf{R}}_1$ and/or $\bar{\mathbf{R}}_2$ elliptic, with $\bar{\mathbf{R}}_2 \neq -\bar{\mathbf{R}}_1$ | $m_{00}, D$ | $p_1^2, p_2^2, P_S$ |
| Single retarder | $\bar{\mathbf{R}}_1$ elliptic, and $\bar{\mathbf{R}}_2 = -\bar{\mathbf{R}}_1$ | $m_{00}, D, \eta, \Delta$ | $p_1^2, p_2^2, P_S$ |
| Dual rotation | $\bar{\mathbf{R}}_1$ and $\bar{\mathbf{R}}_2$ circular, with $\bar{\mathbf{R}}_2 \neq -\bar{\mathbf{R}}_1$ | $m_{00}, D, \bar{R}_C,$ $D_C, P_C$ | $p_1^2, p_2^2, P_S, \bar{R}$ $\bar{R}_L, D_L, P_L, \Delta$ |
| Single rotation | $\bar{\mathbf{R}}_1$ and $\bar{\mathbf{R}}_2$ circular, with $\bar{\mathbf{R}}_2 = -\bar{\mathbf{R}}_1$ | $m_{00}, D, \bar{R}_C,$ $D_C, P_C, \eta$ | $p_1^2, p_2^2, P_S, \bar{R}$ $\bar{R}_L, D_L, P_L, \Delta$ |

Central arguments of polarization algebra and polarimetry like the facts that any Mueller matrix can be considered as a convex sum of pure Mueller matrices, that the polarimetric behavior of many media is essentially nondepolarizing, and that any depolarizing Mueller matrix has an associated pure one, justify the interest of the approach presented in this work. It provides a complementary view of the information contained in pure Mueller matrices by means of meaningful physical quantities.

**Funding Information.** Ministerio de Economía y Competitividad (FIS2014-58303-P); Gobierno de Aragón (E99).